# Acceleration toward polarization singularity inspired by relativistic E×B drift

Sunkyu Yu, Xianji Piao, and Namkyoo Park*

*Photonic Systems Laboratory, Department of Electrical and Computer Engineering, Seoul National University, Seoul 08826, Korea*

*E-mail address for correspondence: nkpark@snu.ac.kr

**The relativistic trajectory of a charged particle driven by the Lorentz force is different from the classical one, by velocity-dependent relativistic acceleration term. Here we show that the evolution of optical polarization states near the polarization singularity can be described in analogy to the relativistic dynamics of charged particles. A phase transition in parity-time symmetric potentials is then interpreted in terms of the competition between electric and magnetic 'pseudo'-fields applied to polarization states. Based on this Lorentz pseudo-force representation, we reveal that zero Lorentz pseudo-force is the origin of recently reported strong polarization convergence to the singular state at the exceptional point. We also demonstrate the deterministic design of achiral and directional eigenstates at the exceptional point, allowing an anomalous linear polarizer which operates orthogonal to forward and backward waves. Our results linking parity-time symmetry and relativistic electrodynamics show that previous PT-symmetric potentials for the polarization singularity with a chiral eigenstate are the subset of optical potentials for the $E{\times}B$ "polarization" drift.**

**Introduction**

With the universal existence of open systems[1,2] of non-equilibrium and time-dependent[3-5] potential energy, the concept of parity-time (PT) symmetry[6,7] has become a multidisciplinary topic[8-15]. PT symmetry successfully offers the special form of potentials $V(x) = V^*(-x)$, providing physical observables even for non-equilibrium systems. The existence of real observables in PT-symmetric complex potentials has opened the field of non-Hermitian quantum mechanics[6,8], which exhibits the phase transition[6,16,17] between real and complex eigenspectra in stark contrast to a purely real eigenspectrum observed in Hermitian potentials. By utilizing optical gain- and loss- materials in the refractive index form $n(x) = n^*(-x)$, the physics of PT-symmetric potentials has been applied to the exotic control of light flows[17,18]. The effective realization of PT symmetry has also been extended to acoustics[9,13,19], optomechanics[10], electronics[11], gyrotropic systems[20,21], and population genetics[12]. For all of these fields, critical traits of PT symmetry, e.g. unidirectionality[22-24], non-Hermitian degeneracy[18], and chirality[25,26], impose intriguing features on wave dynamics in terms of the 'singularity'[27,28]: the coalescence of eigenstates with a chiral form[25,26,29-32] at the exceptional point (EP, or phase transition point)[33].

Meanwhile, it is known that relativistic electrodynamics[34] for charged particles also exhibits the inherent feature of open systems. The famous relativistic energy expression[35], $\tilde{E} = mc^2 / [1 - (v/c)^2]^{1/2}$, shows that observers in different frames will see different values of total energy for moving charged particles of velocity $v$. We note that this non-equilibrium condition results in non-Hermitian form of Hamiltonians, the necessary condition for the achievement of PT symmetry. In the context of the multidisciplinary realization of PT symmetry[8-12], therefore, the link between relativistic behaviors of charged particles in electromagnetic fields and wave dynamics in PT-symmetric potentials could offer different viewpoints on the physics of EP singularity in PT-symmetric potentials.

Inspired by the polarization equation of motion from the Schrödinger-like form of Maxwell's equations[36], here we interpret the evolution of optical states of polarizations (SOP) near the EP singularity in direct analogy to the relativistic $E×B$ drift (the movement under orthogonal $E$ and $B$ fields) of charged particles[34], which we call the relativistic $E×B$ "polarization" drift of light. The phase transition in PT-symmetric potentials[6,16,17] is then understood in view of the competition between electric- and magnetic- 'pseudo'-fields, and we prove that strong chiral conversion of optical SOP at the EP[15,16] corresponds to the accidental cancellation of the Lorentz pseudo-force on the Poincaré hemisphere. By employing this "Lorentz-force picture" in the analysis of the polarization singularity, we then extend the class of the polarization singularity in vector wave equations[25,26,37-39], revealing the existence of achiral and directional eigenstates at the EP. Our approach paves the way for the unconventional control of optical polarizations, such as anomalous directional polarizers.

**Results**

**Lorentz pseudo-forces for optical polarizations**   Consider the planewave propagating along the $z$-axis of the electrically anisotropic material, with the unity permeability ($\mu = 1$). For the later discussion, we express the arbitrary permittivity tensor in the $x$-$y$ plane, in terms of Pauli matrices[26,31] $\sigma_{1-3}$ as $\boldsymbol{\varepsilon} = (\varepsilon_2\sigma_1 + \varepsilon_3\sigma_2 + \varepsilon_1\sigma_3)$, or

$$\boldsymbol{\varepsilon} = \begin{pmatrix} \varepsilon_o + \varepsilon_1 & \varepsilon_2 - i\varepsilon_3 \\ \varepsilon_2 + i\varepsilon_3 & \varepsilon_o - \varepsilon_1 \end{pmatrix}, \quad (1)$$

where $\varepsilon_o(z)$ and $\varepsilon_{1-3}(z)$ have slowly-varying complex values, $\sigma_1 = [0,1;1,0]$, $\sigma_2 = [0,-i;i,0]$, and $\sigma_3 = [1,0;0,-1]$. By applying the spin $(\mathbf{x} \pm i\mathbf{y})/2^{1/2}$ bases[36], Maxwell's equations become the vector Schrödinger-like equation $d\boldsymbol{\psi_e}/dz = H_s \cdot \boldsymbol{\psi_e}$ with the temporal-like $z$-axis[40], where the spinor representation of $\boldsymbol{\psi_e} = [\psi_{e+}, \psi_{e-}]^T$ is the electric field amplitude of ($\pm$) optical spin waves ((+) for right-circular polarization (RCP) $(\mathbf{x} + i\mathbf{y})/2^{1/2}$, and (−) for left-circular polarization (LCP) $(\mathbf{x} - i\mathbf{y})/2^{1/2}$), and $H_s$ is the traceless Hamiltonian expressed with Pauli matrices[36,41] as $H_s = (\varepsilon_1\sigma_1 + \varepsilon_2\sigma_2 + \varepsilon_3\sigma_3)/(i\lambda)$ for $\lambda = 2\varepsilon_o^{1/2}/k_0$ and the free-space wavenumber $k_0$. If we assign the symmetry axis between $x$ and $y$ axes, the condition of PT-symmetric potentials[15,16] requires real-valued $\varepsilon_o$, $\varepsilon_2$, and $\varepsilon_3$, and imaginary-valued $\varepsilon_1$. Note that imaginary-valued $\varepsilon_1$ corresponds to the linear dichroism[42,43], the different dissipation for each linear polarization, while real-valued $\varepsilon_2$ represents the birefringence. $\varepsilon_3$ represents the magneto-optical change of plasma permittivity induced by an external static magnetic field[34], and for now, we assume the nonmagnetic case of $\varepsilon_3 = 0$.

In this representation, the SOP of light is described by the Stokes parameters[44] $S_j = \psi_e^\dagger \cdot \sigma_j \cdot \psi_e$ ($j = 0,1,2,3$). The change of the SOP can then be expressed in view of light-matter interactions[36] by applying the governing equation $d\psi_e/dz = H_s \cdot \psi_e$ and its conjugate form $d\psi_e^\dagger/dz = \psi_e^\dagger \cdot H_s^\dagger$, which leads to the Lorentz pseudo-force equation of motion for the SOP[36] (also see Supplementary Note 1)

$$\frac{d}{dz}\mathbf{S_n} = \mathbf{E} + \mathbf{S_n} \times \mathbf{B} - (\mathbf{S_n} \cdot \mathbf{E})\mathbf{S_n}, \qquad (2)$$

where $\mathbf{S_n} = [S_1, S_2, S_3]^T/S_0$ is the pseudo-velocity of the 'hypothetical' charged particle corresponding to the SOP of light, and $\mathbf{E} = 2 \cdot Im[\varepsilon_1, \varepsilon_2, \varepsilon_3]^T/\lambda$ and $\mathbf{B} = -2 \cdot Re[\varepsilon_1, \varepsilon_2, \varepsilon_3]^T/\lambda$ are the electric and magnetic pseudo-field in relation to imaginary- and real-parts of the permittivity, respectively. Note that Eq. (2) provides direct analogy to the relativistic dynamics of massless charged particles with the motion equation[20,22] of $\partial_t \boldsymbol{\beta} = \mathbf{E} + \boldsymbol{\beta} \times \mathbf{B} - (\boldsymbol{\beta} \cdot \mathbf{E})\boldsymbol{\beta}$. In this representation of optical polarization states, the acceleration of optical SOP comes from the Lorentz pseudo-force, $\mathbf{F} \sim d\mathbf{S_n}/dz$ (Fig. 1). The first two terms of Eq. (2) are the counterparts of the classical electromagnetic Lorentz force $\mathbf{E} + \boldsymbol{\beta} \times \mathbf{B}$, and the 3rd term $-(\mathbf{S_n} \cdot \mathbf{E})\mathbf{S_n}$ corresponds to the Joule effect[34] $-(\boldsymbol{\beta} \cdot \mathbf{E})\boldsymbol{\beta}$ in the relativistic equation of motion.

Figure 1a-c shows the effect of each component of the Lorentz pseudo-force on the SOP of propagating light, induced by nonmagnetic PT-symmetric materials (imaginary $\varepsilon_1$, real $\varepsilon_2$, and zero $\varepsilon_3$). While $\varepsilon_2$ of the birefringence derives the circulating acceleration on the Poincaré sphere ($\mathbf{B} = -2\varepsilon_2 \cdot \mathbf{e_2}/\lambda$, Fig. 1a, Hermitian case), $\varepsilon_1$ of the amplification or dissipation results in the linear drift of the SOP ($\mathbf{E} = 2 \cdot Im[\varepsilon_1] \cdot \mathbf{e_1}/\lambda$, Fig. 1b, non-Hermitian case). We also note that the energy variation from gain and loss materials $\mathbf{S_n} \cdot \mathbf{E}$ (Fig. 1c) provides the relativistic nonlinear acceleration of the SOP with respect to $\mathbf{E}$.

Consequently, with the orthogonality between pseudo-fields (**E**⊥**B**), PT-symmetric potentials naturally satisfy the ideal $E×B$ drift[34,45] condition to optical polarization states.

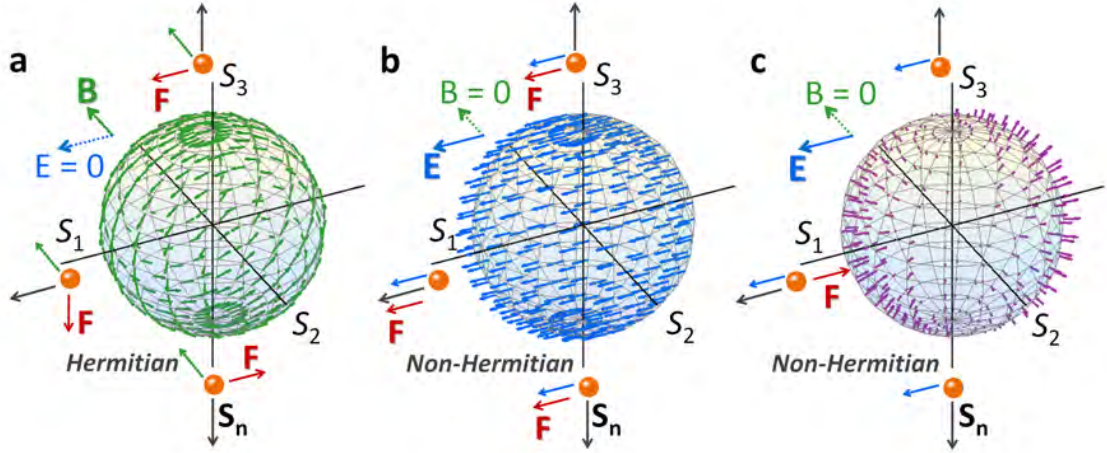

**Figure. 1. Lorentz pseudo-forces for the acceleration of optical SOP.** Classical (**a**) magnetic ($S_n×B$) and (**b**) electric accelerations (**E**) on the Poincaré sphere. (**c**) The acceleration from the relativistic energy variation ($-S_n(S_n·E)$), originating from the PT-symmetric pseudo-Hermiticity of the Hamiltonian $H_s$. The corresponding forces **F** (red arrows) for accelerating positively-charged pseudo-particles (orange spheres) with different pseudo-velocities (or different SOPs, black arrows of $S_n$) are also shown in (**a-c**), for $S_n = e_1$, $e_3$, and $-e_3$.

**Lorentz force picture on PT symmetry** Based on the Lorentz pseudo-force equation of Eq. (2), a phase transition[6,16,17] between real and complex eigenspectra in PT-symmetric potentials can be interpreted in terms of the $E×B$ drift[34,45]: the competition between electric and magnetic pseudo-forces. Figure 2a,b shows the evolution of real and imaginary eigenvalues $\Delta\varepsilon_{eig}$ for the Hamiltonian equation $d\boldsymbol{\psi}_e/dz = H_s·\boldsymbol{\psi}_e$, as a function of the imaginary potential ($\varepsilon_i = Im[\varepsilon_1]$, $\varepsilon_c = \varepsilon_2$). First, before the EP where eigenvalues are real and non-degenerate (Fig. 2a, $\varepsilon_i < \varepsilon_c$), the magnetic pseudo-field is larger than the electric pseudo-field, resulting in the counter-directive acceleration of SOP (lower panels in Fig. 2c) to northern-/southern-hemispheres. At the EP with the coalescence (*d* point in Fig. 2a,b, $\varepsilon_i = \varepsilon_c$), the equal magnitude of $\mathbf{E} = 2·\varepsilon_i·\mathbf{e_1}/\lambda$ and $\mathbf{B} = -2\varepsilon_c·\mathbf{e_2}/\lambda$ fields derives the suppression of total Lorentz pseudo-force on the southern Poincaré sphere, especially with the zero net force at the south pole ($S_n = -e_3$, $dS_n/dz =$ $\mathbf{E} + \mathbf{S_n} × \mathbf{B} - (\mathbf{S_n}·\mathbf{E})\mathbf{S_n} = 2·\varepsilon_i·\mathbf{e_1}/\lambda + \mathbf{e_3} × 2\varepsilon_c·\mathbf{e_2}/\lambda = 0$, Fig. 2d). It is emphasized that this force cancellation impedes the acceleration near the south pole of the stationary polarization, deriving the SOP convergence to perfect LCP chirality[26]. After the EP with amplifying and dissipative states (Fig. 2b, $\varepsilon_i > \varepsilon_c$), the strong electric pseudo-field dominates the motion equation of the SOP, with the co-directive force (lower panels in Fig. 2e) to opposite hemispheres. In the context of electrodynamics analogy, the phase of eigenvalues in PT-symmetric potentials can thus be divided by the (i) **B**-dominant (before the EP), (ii) **B** = **E** (at the EP) and (iii) **E**-dominant regime (after the EP). It is worth mentioning that the stable point with the stationary polarization can also be obtained at the north pole by changing the sign of $\varepsilon_c$ (converting the fast and slow axes for the birefringence) or $\varepsilon_i$ (converting the gain and loss axes for the linear dichroism), allowing perfect RCP chirality. In terms of this Lorentz pseudo-force representation of SOP, we also note that PT-symmetric potentials[15,16] with real-valued $\varepsilon_2$ and imaginary-valued $\varepsilon_1$ are the special case of the $E \times B$ drift with specific field vectors $\mathbf{E} = 2 \cdot \varepsilon_i \cdot \mathbf{e_1}/\lambda$ and $\mathbf{B} = -2\varepsilon_c \cdot \mathbf{e_2}/\lambda$, implying the existence of unconventional polarization singularity at other SOPs (e.g. without optical spin) which will be discussed later.

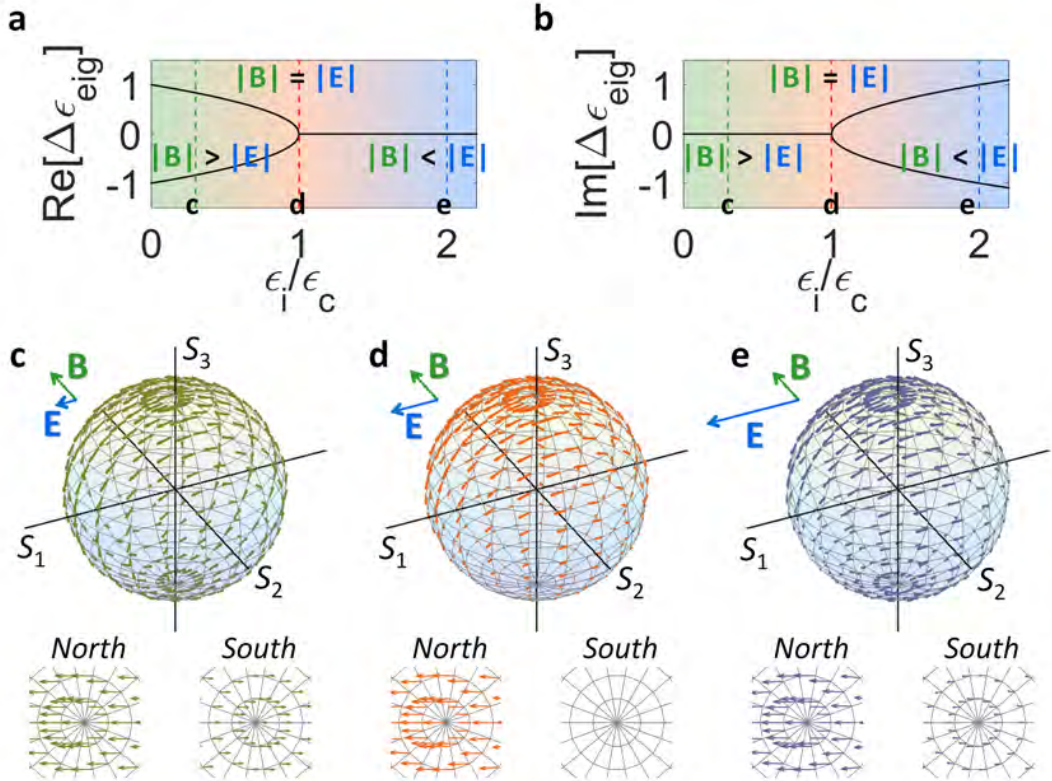

**Figure 2. Phases of PT symmetry in terms of the $E \times B$ drift.** The evolution of the eigenvalues $\Delta\varepsilon_{eig}$ is shown in (**a**) for their real parts ($Re[\Delta\varepsilon_{eig}]$), and in (**b**) for imaginary parts ($Im[\Delta\varepsilon_{eig}]$). The Lorentz pseudo-force acceleration

for each phase of PT symmetry is shown: **(c)** before the EP with $|\mathbf{B}| > |\mathbf{E}|$, **(d)** at the EP with $|\mathbf{B}| = |\mathbf{E}|$, and **(e)** after the EP with $|\mathbf{B}| < |\mathbf{E}|$. The enlarged plots in **(c-e)** show the distribution of accelerations near the north and south poles.

***E×B* polarization drift in PT-symmetric potentials***   We then investigate the "evolution" of the SOP, for different pseudo-forces shown in Fig. 2. Figure 3a-c shows the change of the initial SOP of (+, RCP) and (−, LCP) spins under different phases of PT symmetry, in relation to the charged particle movement at different phases[45] of the relativistic *E×B* drift (Fig. 3d-3f). Because the magnetic pseudo-field dominates the dynamics of SOP before the EP (Fig. 3a,d, $|\mathbf{B}| > |\mathbf{E}|$), the SOP for each spin simply rotates around the **B** field following the $\mathbf{S_n} \times \mathbf{B}$ of Eq. (2). Yet, with different magnitudes of the forces in northern and southern hemispheres (Fig. 2c), the 'speed' of the SOP rotation near each pole is different, analogous to different magnetically-gyrating arcs of charged particles in the *E×B* drift[34,45]. The directional drift of relativistic particles along the **E×B** axis (Fig. 3d, toward $-S_3$ axis) is thus reproduced by the slow evolution of SOPs near the $\mathbf{S_n} = -\mathbf{e_3}$ on the Poincaré sphere (Fig. 3a).

The extraordinary case of the relativistic *E×B* polarization drift is achieved at the singular state of EP (Fig. 3b,e), for the case of $|\mathbf{B}| = |\mathbf{E}|$. Because of the force cancellation near the perfectly stable south pole ($\mathbf{S_n} = -\mathbf{e_3}$ for $d\mathbf{S_n}/dz = 0$), the (+) spin state converges to the (–) spin when the state approaches the south pole through the gyration by the magnetic pseudo-field (orange line in Fig. 3b), similar to the convergence of the velocity in the motion of relativistic particles (orange line in Fig. 3e). Because the (−) spin state is stationary, we note that the entire SOP, which can be represented in terms of the linear combination of the LCP (− spin) and RCP (+ spin), is thus converted to the LCP chiral wave. After the EP ($|\mathbf{B}| < |\mathbf{E}|$, Fig. 3c,f), the electric force is dominant, resulting in the linear acceleration mostly towards the direction of **E**. For all cases, it is noted that the relativistic correction from non-Hermitian Hamiltonians retains the evolution of SOP $\mathbf{S_n} = [S_1, S_2, S_3]^T / S_0$ on the Poincaré sphere, in contrast to the classical evolutions (dotted lines in Fig. 3a-c) which do not include the third term of Eq. (2).

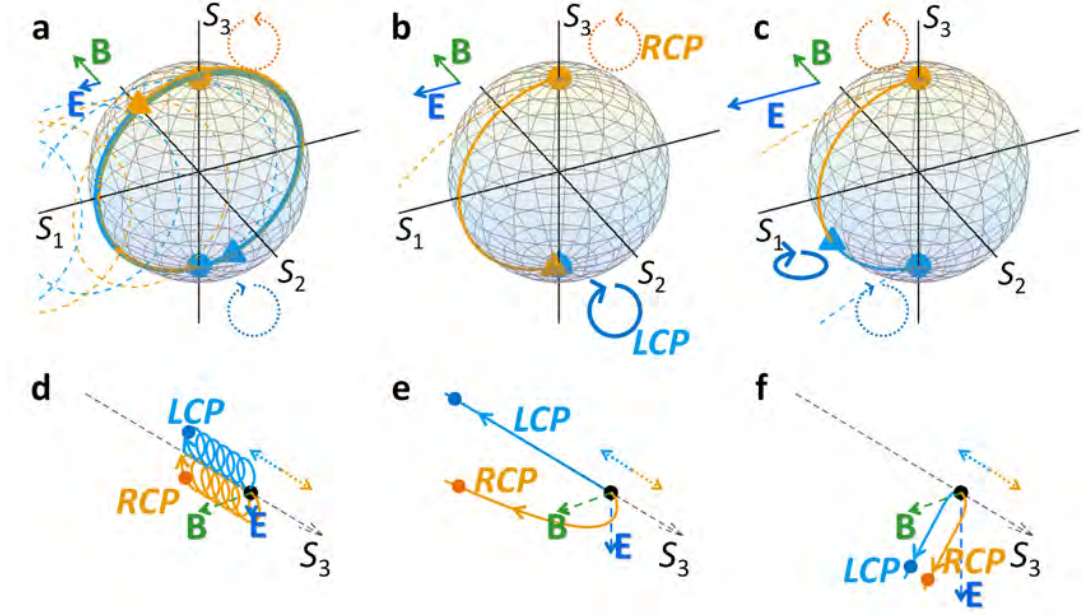

**Figure 3. Evolutions of optical SOP in PT-symmetric potentials. (a-c)** The change of the SOP at different phases of PT symmetry: **(a)** before the EP with $|\mathbf{B}| > |\mathbf{E}|$, **(b)** at the EP with $|\mathbf{B}| = |\mathbf{E}|$, and **(c)** after the EP with $|\mathbf{B}| < |\mathbf{E}|$. The corresponding movements of charged particles by the $E \times B$ drift are shown in **(d-f)**, respectively. While circles and triangles each denote the incident and final (after $10\varepsilon_o^{1/2} \cdot \lambda_o / \pi$) state, orange color is for the initial RCP (or positive spin, $\mathbf{S_n} = \mathbf{e_3}$), and blue color is for the initial LCP (or negative spin, $\mathbf{S_n} = -\mathbf{e_3}$). The dotted lines in **(a-c)** represent the non-relativistic movements of the SOP, without the third term in Eq. (2). The dotted arrows for the LCP and RCP in **(d-f)** denote the initial movement of each spin.

**Realization of achiral and directional singularity**  Extending the special case of the $E \times B$ polarization drift derived from $\mathbf{E} = 2 \cdot \varepsilon_i \cdot \mathbf{e_1}/\lambda$ and $\mathbf{B} = -2\varepsilon_c \cdot \mathbf{e_2}/\lambda$, we now work on other types of $E \times B$ polarization drift which allow unconventional polarization singularity without optical spin, by manipulating the direction of electromagnetic pseudo-fields. The vector form of the Lorentz pseudo-force equation provides larger degrees of freedom for the intuitive control of the eigenstate at the singularity, in contrast to the fixed chiral form[25,26,29-32] of scalar PT-symmetric equation. Although the magnetic transition of PT symmetry can also be utilized to achieve the achiral (spin-less) eigenstate at the EP ($\varepsilon_3 \neq 0$, Supplementary Note 2), here we investigate the realization of achiral and directional singularity with the use of nonmagnetic chiral materials. From the constitutive relation including optical chirality[46] (or bi-isotropy) $\mathbf{D} = \varepsilon\mathbf{E} - i\chi\mathbf{H}$ and $\mathbf{B} = \mu\mathbf{H} + i\chi\mathbf{E}$, the condition of $\boldsymbol{\mu} = \mu_0$ and diagonal $\boldsymbol{\varepsilon}$ with $\varepsilon_x \neq \varepsilon_y$ represents the nonmagnetic chiral material with general electrical anisotropy, including both birefringence ($Re[\varepsilon_x] \neq Re[\varepsilon_y]$ for different $x$- and $y$-wavevector) and linear dichroism[42,43] ($Im[\varepsilon_x] \neq Im[\varepsilon_y]$ for different $x$- and $y$-dissipation). The spin-based Hamiltonian equation $d\psi_e/dz = H_s \cdot \psi_e$, for slowly-varying $\varepsilon_x = \varepsilon_o + \Delta\varepsilon(z)$ and $\varepsilon_y = \varepsilon_o - \Delta\varepsilon(z)$ and constant $\chi = \chi_o$ with real-valued $\varepsilon_o$ and $\chi_o$, derives the Hamiltonian $H_s$ for general chiral materials, in the form of

$$H_s = \begin{bmatrix} i\omega\chi & -\frac{i}{2} \cdot (k - \omega\chi) \cdot \left(\frac{\Delta\varepsilon}{\varepsilon_o}\right) \\ -\frac{i}{2} \cdot (k + \omega\chi) \cdot \left(\frac{\Delta\varepsilon}{\varepsilon_o}\right) & -i\omega\chi \end{bmatrix}, \quad (3)$$

and $k^2 = \omega^2 \cdot (\mu_0 \varepsilon_o + \chi_o^2)$ (see Supplementary Note 3 for the general case of spatially-varying $\chi(z)$). The Pauli expression[41] of $H_s = a_1\sigma_1 + a_2\sigma_2 + a_3\sigma_3$ has the coefficients of $a_1 = -(ik/2) \cdot (\Delta\varepsilon/\varepsilon_o)$, $a_2 = -(\omega\chi/2) \cdot (\Delta\varepsilon/\varepsilon_o)$, and $a_3 = i\omega\chi$. The electric and magnetic pseudo-fields for Eq. (2) are then defined as

$$\mathbf{E} = \begin{bmatrix} k \cdot Im(\rho) \\ -\omega\chi_o \cdot Re(\rho) \\ 0 \end{bmatrix}, \quad \mathbf{B} = \begin{bmatrix} -k \cdot Re(\rho) \\ -\omega\chi_o \cdot Im(\rho) \\ 2\omega\chi_o \end{bmatrix}, \quad (4)$$

with the degree of electrical anisotropy $\rho = \Delta\varepsilon/\varepsilon_o$.

Equation (4) proves that the pseudo-field components $\mathbf{E}(\rho,\chi)$ and $\mathbf{B}(\rho,\chi)$ driving the SOP are strongly dependent on the type of the anisotropy: birefringence (real $\rho$) or linear dichroism (imaginary $\rho$) both satisfying the condition of the $E \times B$ drift ($\mathbf{E} \perp \mathbf{B}$). For the case of birefringence with $\mathbf{E}(\rho,\chi) = -\omega\chi_o\rho \cdot \mathbf{e_2}$ and $\mathbf{B}(\rho,\chi) = -k\rho \cdot \mathbf{e_1} + 2\omega\chi_o \cdot \mathbf{e_3}$, the pseudo-field satisfies $|\mathbf{E}| < |\mathbf{B}|$ in most cases and the condition of $|\mathbf{E}| \geq |\mathbf{B}|$ enforces $\varepsilon_o < -4\chi_o^2/(\mu_0 \cdot \rho^2)$ and thus prohibits the existence of propagating waves at the EP. It is interesting to note that this restriction proves the necessity of complex potentials for obtaining the singularity; therefore we employ linear dichroism for achieving the EP for the propagating wave, by fulfilling the condition of $|\mathbf{B}| = |\mathbf{E}|$.

Figure 4 shows the case of linearly-dichroic ($\rho = i \cdot \rho_i$) chiral materials, which derive pseudo-fields of $\mathbf{E}(\rho,\chi) = k\rho_i \cdot \mathbf{e_1}$ and $\mathbf{B}(\rho,\chi) = -\omega\chi_o\rho_i \cdot \mathbf{e_2} + 2\omega\chi_o \cdot \mathbf{e_3}$ with the EP condition of $\rho_i^2 = 4\chi_o^2/(\mu_0\varepsilon_o)$ for $|\mathbf{B}| = |\mathbf{E}|$.

The PT-symmetry-like phase transition (Fig. 4b-d) around the EP (marked with red dots in Fig. 4c,f) occurs in linearly-dichroic chiral materials from the competition between **E** and **B** (Fig. 4a), and the direction of the *E*×*B* drift is controlled by changing $\rho_i$ for the pseudo-magnetic field **B**, allowing the realization of the achiral singularity ($\mathbf{S_n} \sim -\mathbf{e_2}$ in Fig. 4c); in sharp contrast to the case of PT-symmetric potentials. Furthermore, the obtained EP state has the directionality in its propagation due to the wavevector-dependency of Eq. (4) (Fig. 4b-d *vs* Fig. 4e-g, $\mathbf{S_n} \sim \mathbf{e_2}$ in Fig. 4f), which originates from the broken mirror symmetry of chiral materials for forward and backward waves.

This directionality with achiral designer eigenstate at the EP allows the implementation of unconventional polarizers based on the polarization convergence at the EP. Figure 4h shows an example of the anomalous linear polarizer, operating 'orthogonal' to forward and backward waves. While the SOP of forward waves are converged to the +45° linear polarization (Fig. 4c), the SOP of backward waves becomes −45° linear polarization. Moreover, in contrast to the case of classical linear polarizers which perfectly reflect the orthogonally polarized waves (e.g. the *y*-polarized incidence to the *x*-polarizer), the linear-polarizing functionality in the structure of Fig. 4h operates for the 'entire' SOP due to the non-orthogonality between eigenstates.

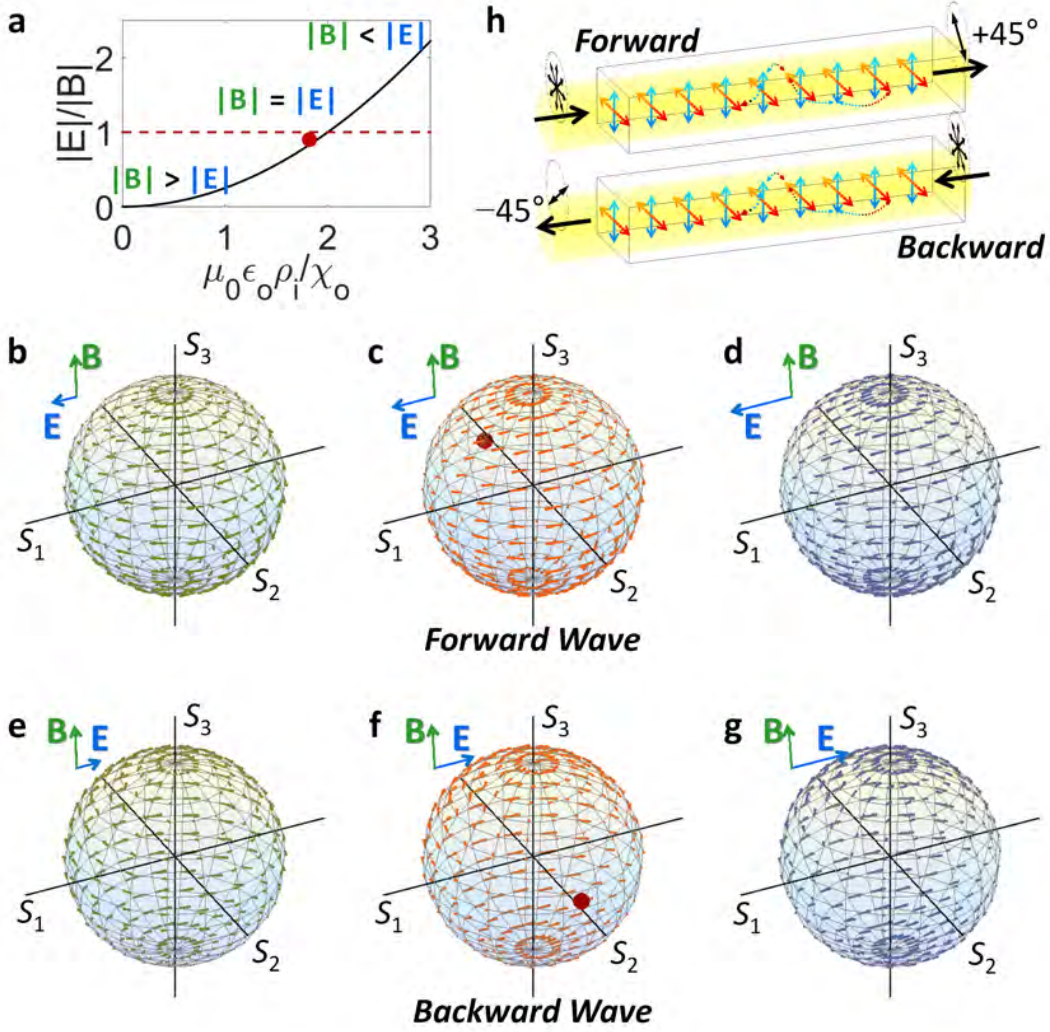

**Figure 4. Directional EP with the achiral eigenstate in linearly-dichroic chiral materials.** (**a**) The relative magnitude of electric and magnetic pseudo-fields as a function of material parameters. The Lorentz pseudo-force acceleration for each phase is shown: (**b,e**) before the EP with $|\mathbf{B}| > |\mathbf{E}|$, (**c,f**) at the EP with $|\mathbf{B}| = |\mathbf{E}|$, and (**d,g**) after the EP with $|\mathbf{B}| < |\mathbf{E}|$, for (**b-d**) forward and (**e-g**) backward waves. (**h**) The operation schematic of an anomalous linear polarizer for randomly-polarized incidences: red and blue arrows denote the anisotropic permittivity with the linear dichroism and dotted arrows represent the coupling through optical chirality.

## Discussion

In summary, we found the link between the seemingly unrelated fields of PT symmetry optics and relativistic electrodynamics. This reinterpretation of PT symmetry brings insight to the singularity in polarization space, broadening the class of parity-time symmetric Hamiltonians in vector wave equations[39]. The counterintuitive achievement of the achiral and directional EP eigenstate is also demonstrated, which allows the realization of anomalous linear polarizers for randomly polarized incidences. The comprehensible understanding of the EP in terms of the dynamics of charged particles will provide a novel design methodology near the singularity: the generation of chiral waves[47-49], topological photonics which has focused only on chiral states[38], coherent wave dynamics[50], PT-symmetry-like potentials based on causality[51,52] or supersymmetric optics[53,54], and optical analogy of spintronics. At the same time, this classical viewpoint on relativistic electrodynamics also enables the analogy of EP dynamics in charged particle movements, the linear $E \times B$ drift toward a single direction for every initial velocity vectors.

**Supplementary Information** is available in the online version of this paper.

**Acknowledgments**

This work was supported by the National Research Foundation of Korea (NRF) through the Global Frontier Program (GFP, 2014M3A6B3063708) and the Global Research Laboratory Program (GRL, K20815000003), all funded by the Ministry of Science, ICT & Future Planning of the Korean government. S. Yu was also supported by the Basic Science Research Program (2016R1A6A3A04009723), and X. Piao and N. Park were also supported by the Korea Research Fellowship Program (KRF, 2016H1D3A1938069) through the NRF, all funded by the Ministry of Education of the Korean government.

**Author Contributions**

S.Y. conceived the presented idea. S.Y. and X.P. developed the theory and performed the computations. N.P. encouraged S.Y. to investigate the link between relativistic electrodynamics and non-Hermitian physics while supervising the findings of this work. All authors discussed the results and contributed to the final manuscript.

**Competing Interests Statement**

The authors declare that they have no competing financial interests.13

# Acceleration toward polarization singularity inspired by relativistic E×B drift

Sunkyu Yu, Xianji Piao, and Namkyoo Park[*]

*Photonic Systems Laboratory, Dept. of Electrical and Computer Engineering, Seoul National University, Seoul 08826, Korea*

**Supplemental Materials**

**Supplementary Note 1. The equation of motion for the state of polarization**

Starting from the spin-based Hamiltonian equation $d\psi_e/dz = H_s \cdot \psi_e$, the equation of motion for the state of polarization (SOP) [1] can be derived. The Stokes parameters [2] $S_j$ ($j = 0,1,2,3$) which quantify the SOP can be decomposed with the inclusion of the Pauli matrices, as $S_j = \psi_e^\dagger \cdot \sigma_j \cdot \psi_e$. The derivatives of $S_j$ then become

$$\frac{dS_j}{dz} = \frac{d}{dz}\left(\vec{\psi}_e^{\,\dagger} \cdot \sigma_j \cdot \vec{\psi}_e\right) = \frac{d\vec{\psi}_e^{\,\dagger}}{dz} \cdot \sigma_j \vec{\psi}_e + \vec{\psi}_e^{\,\dagger} \sigma_j \cdot \frac{d\vec{\psi}_e}{dz}. \tag{S1}$$

By applying $d\psi_e/dz = H_s \cdot \psi_e$ and $d\psi_e^\dagger/dz = \psi_e^\dagger \cdot H_s^\dagger$ to Eq. (S1), the 'SOP change' operator can be introduced as $S_{d,j} = H_s^\dagger \cdot \sigma_j + \sigma_j \cdot H_s$, from the relation of

$$\frac{dS_j}{dz} = \vec{\psi}_e^{\,\dagger} \cdot \left(H_s^\dagger \sigma_j + \sigma_j H_s\right) \cdot \vec{\psi}_e = \vec{\psi}_e^{\,\dagger} \cdot \hat{S}_{d,j} \cdot \vec{\psi}_e, \tag{S2}$$

where the expectation value of the operator $S_{d,j}$ directly represents the change of each Stokes parameter $dS_j/dz$.

The Pauli notation of the Hamiltonian $H_s = a_0\sigma_0 + a_1\sigma_1 + a_2\sigma_2 + a_3\sigma_3$ derives the intuitive expression of the SOP change operator as

$$\begin{aligned}
\hat{S}_{d,0} &= H_s^\dagger \sigma_0 + \sigma_0 H_s = 2 \cdot \sum_{j=0}^{3} \mathrm{Re}[a_j] \cdot \sigma_j \\
\hat{S}_{d,p} &= H_s^\dagger \sigma_p + \sigma_p H_s \\
&= 2 \cdot \mathrm{Re}[a_0] \cdot \sigma_p + 2 \cdot \mathrm{Re}[a_p] \cdot \sigma_0 - 2 \cdot \{\mathrm{Im}[a_q] \cdot \sigma_r - \mathrm{Im}[a_r] \cdot \sigma_q\}
\end{aligned} \tag{S3}$$

where $p = 1,2,3$, and $(p,q,r)$ is the cyclic order of $(1,2,3)$. The expectation values of Eq. (S3) define the change of $S_j$ in terms of the Stokes parameters as

$$\begin{aligned}
\frac{dS_0}{dz} &= 2 \cdot \sum_{j=0}^{3} \mathrm{Re}[a_j] \cdot S_j = 2 \cdot \left(\mathrm{Re}[a_0] \cdot S_0 + \sum_{p=1}^{3} \mathrm{Re}[a_p] \cdot S_p\right) \\
\frac{dS_p}{dz} &= 2 \cdot \mathrm{Re}[a_0] \cdot S_p + 2 \cdot \mathrm{Re}[a_p] \cdot S_0 - 2 \cdot \{\mathrm{Im}[a_q] \cdot S_r - \mathrm{Im}[a_r] \cdot S_q\}
\end{aligned} \tag{S4}$$

Equation (S4) can be expressed in the form similar to the governing equation of electrodynamics [3] as

$$\frac{dS_0}{dz} = 2 \cdot \text{Re}[a_0] \cdot S_0 + \mathbf{S} \cdot \mathbf{E}$$
$$\frac{d\mathbf{S}}{dz} = 2 \cdot \text{Re}[a_0] \cdot \mathbf{S} + S_0 \cdot \mathbf{E} + \mathbf{S} \times \mathbf{B}$$
(S5)

where $\mathbf{S} = [S_1, S_2, S_3]^T$ is the Stokes vector, and $\mathbf{E} = 2 \cdot Re[a_1, a_2, a_3]^T$ and $\mathbf{B} = 2 \cdot Im[a_1, a_2, a_3]^T$ each represents the 'pseudo-' electric and magnetic field which drives the Lorentz-like force to the SOP. The condition of $a_0 = 0$ then transforms Eq. (S5) into the simplified form of

$$\frac{dS_0}{dz} = \mathbf{S} \cdot \mathbf{E}$$
$$\frac{d\mathbf{S}}{dz} = S_0 \cdot \mathbf{E} + \mathbf{S} \times \mathbf{B}$$
(S6)

The expressions of the pseudo-fields $\mathbf{E}$ and $\mathbf{B}$ are shown as $\mathbf{E} = 2 \cdot Im[\varepsilon_1, \varepsilon_2, \varepsilon_3]^T$ and $\mathbf{B} = -2 \cdot Re[\varepsilon_1, \varepsilon_2, \varepsilon_3]^T$ for PT-symmetric potentials and as Eq. (4) of the main manuscript for linearly-dichroic chiral materials. By assigning the normalized Stokes parameter [1] as $\mathbf{S_n} = \mathbf{S}/S_0$, Eq. (S6) can also be expressed as $d\mathbf{S_n}/dz = \mathbf{E} + \mathbf{S_n} \times \mathbf{B} - (\mathbf{S_n} \cdot \mathbf{E})\mathbf{S_n}$ (Eq. (2) in the main manuscript).

**Supplementary Note 2. Magnetically-induced transition of EP for the achiral eigenstate**

We introduce the previously neglected term of $\varepsilon_3\sigma_3$, which can be obtained by imposing an external static magnetic field upon plasmas [3] along the $z$-axis. The condition of EP ($|\mathbf{B}| = |\mathbf{E}|$) can be maintained by controlling the relative magnitude of the birefringence ($\varepsilon_2\sigma_2$) and the external magnetic field ($\varepsilon_3\sigma_3$). Counterintuitively [4,5], we can achieve the 'achiral' eigenstate at the EP, which can also be controlled by the strength of the external magnetic field for $\varepsilon_3\sigma_3$ (Fig. S1a-c). This active control of the EP eigenstate lifts the restriction on the phenomena near the EP, which have focused only on the chiral state [5-10].

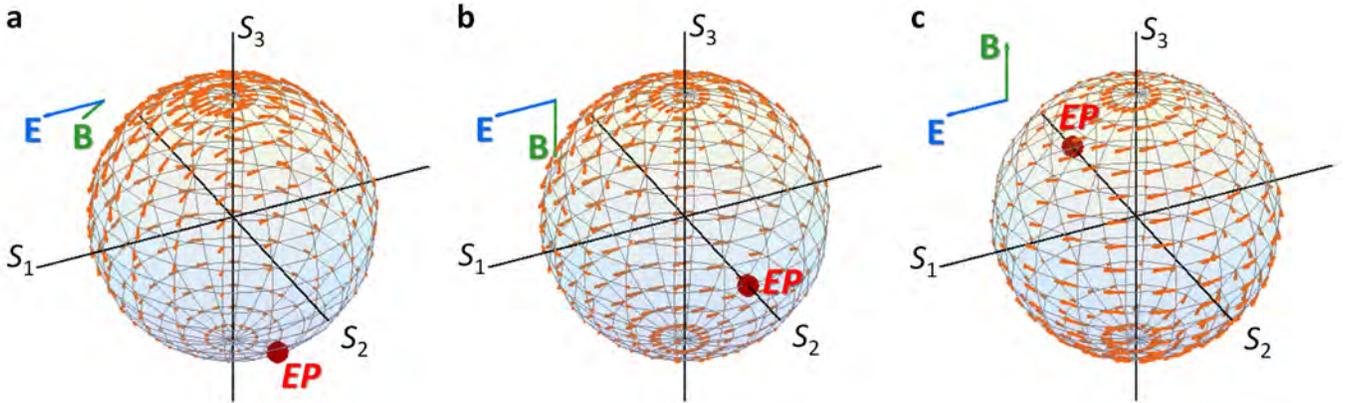

Figure S1. Magnetically-induced transition of EP on the Poincaré sphere. The accelerations at the EP ($|\mathbf{B}| = |\mathbf{E}|$) for different directions of magnetic pseudo-fields: (a) $\mathbf{B} = (\mathbf{e_2} + \mathbf{e_3})/2^{1/2}$, (b) $\mathbf{B} = \mathbf{e_3}$, and (c) $\mathbf{B} = -\mathbf{e_3}$. Red circles denote the EP on the Poincaré sphere, for the SOP of zero Lorentz pseudo-force. The $\mathbf{e_3}$ component of the magnetic pseudo-field originates from the external static magnetic field along the $z$-axis.

**Supplementary Note 3. Derivation of the spin-based Hamiltonian for linearly-dichroic chiral materials**

For the z-axis propagating planewave through the chiral material of $\mu = \mu_0$ and $\varepsilon_x \neq \varepsilon_y$, Maxwell's equations become

$$\frac{\partial E_x}{\partial z} = -i\omega \cdot (\mu_0 H_y + i\chi E_y)$$

$$-\frac{\partial E_y}{\partial z} = -i\omega \cdot (\mu_0 H_x + i\chi E_x)$$

$$\frac{\partial H_x}{\partial z} = i\omega \cdot (\varepsilon_y E_y - i\chi H_y) \quad \text{(S7)}$$

$$-\frac{\partial H_y}{\partial z} = i\omega \cdot (\varepsilon_x E_x - i\chi H_x)$$

Considering the spatially-varying material along the z-axis, each field has the form of $E_{x,y}(z) = \psi_{e(x,y)}(z) \cdot \exp(-ikz)$ and $H_{x,y}(z) = \psi_{h(x,y)}(z) \cdot \exp(-ikz)$ for the constant $k$, and Eq. (S7) is then expressed with the field amplitudes $\psi_{e,h}(z)$ as

$$\psi_{ex}' - ik\psi_{ex} = -i\omega\mu_0 \psi_{hy} + \omega\chi\psi_{ey}$$

$$\psi_{ey}' - ik\psi_{ey} = i\omega\mu_0 \psi_{hx} - \omega\chi\psi_{ex}$$

$$\psi_{hx}' - ik\psi_{hx} = i\omega\varepsilon_y \psi_{ey} + \omega\chi\psi_{hy} \quad \text{(S8)}$$

$$\psi_{hy}' - ik\psi_{hy} = -i\omega\varepsilon_x \psi_{ex} - \omega\chi\psi_{hx}$$

The 4-coupled derivative equations of Eq. (S8) can be transformed to the 2-coupled equations only including the electrical components of $\psi_{e(x,y)}$. By applying the slowly-varying approximation ($|\psi''| \ll |k\psi'|$), Eq. (S8) becomes

$$-2ik \cdot \psi_{ex}' - 2\omega\chi\psi_{ey}' = [k^2 - \omega^2 \cdot (\mu_0\varepsilon_x - \chi^2)] \cdot \psi_{ex} + (\omega\chi' - 2ik\omega\chi) \cdot \psi_{ey}$$

$$2\omega\chi \cdot \psi_{ex}' - 2ik\psi_{ey}' = (-\omega\chi' + 2ik\omega\chi) \cdot \psi_{ex} + [k^2 - \omega^2 \cdot (\mu_0\varepsilon_y - \chi^2)] \cdot \psi_{ey} \quad \text{(S9)}$$

We can now separate the derivative terms for x and y polarizations as

$$2 \cdot (k^2 - \omega^2\chi^2) \cdot \psi_{ex}' = \{ik \cdot [k^2 - \omega^2 \cdot (\mu_0\varepsilon_x + \chi^2)] + \omega^2\chi \cdot \chi'\} \cdot \psi_{ex}$$
$$+ \{\omega\chi \cdot [k^2 + \omega^2 \cdot (\mu_0\varepsilon_y - \chi^2)] + ik\omega \cdot \chi'\} \cdot \psi_{ey}$$

$$2 \cdot (k^2 - \omega^2\chi^2) \cdot \psi_{ey}' = \{-\omega\chi \cdot [k^2 + \omega^2 \cdot (\mu_0\varepsilon_x - \chi^2)] - ik\omega \cdot \chi'\} \cdot \psi_{ex} \quad \text{(S10)}$$
$$+ \{ik \cdot [k^2 - \omega^2 \cdot (\mu_0\varepsilon_y + \chi^2)] + \omega^2\chi \cdot \chi'\} \cdot \psi_{ey}$$

which leads to the governing Hamiltonian equation for the Cartesian basis as $d[\psi_{ex}, \psi_{ey}]^T/dz = H_c \cdot [\psi_{ex}, \psi_{ey}]^T$ where

$$H_c = \frac{1}{2 \cdot (k^2 - \omega^2\chi^2)} \times$$
$$\begin{bmatrix} ik \cdot [k^2 - \omega^2 \cdot (\mu_0\varepsilon_x + \chi^2)] + \omega^2\chi \cdot \chi' & \omega\chi \cdot [k^2 + \omega^2 \cdot (\mu_0\varepsilon_y - \chi^2)] + ik\omega \cdot \chi' \\ -\omega\chi \cdot [k^2 + \omega^2 \cdot (\mu_0\varepsilon_x - \chi^2)] - ik\omega \cdot \chi' & ik \cdot [k^2 - \omega^2 \cdot (\mu_0\varepsilon_y + \chi^2)] + \omega^2\chi \cdot \chi' \end{bmatrix} \quad \text{(S11)}$$

When we apply the spin basis representation using the rotation matrix $M$ as $[\psi_{e+}, \psi_{e-}]^T = M \cdot [\psi_{ex}, \psi_{ey}]^T$ where

$$M = \frac{1}{\sqrt{2}} \begin{bmatrix} 1 & -i \\ 1 & i \end{bmatrix}, \tag{S12}$$

Eq. (S10) is converted to $d\psi_e/dz = d(M[\psi_{ex}, \psi_{ey}]^T)/dz = (MH_cM^{-1}) \cdot (M[\psi_{ex}, \psi_{ey}]^T) = H_s \cdot \psi_e$ for the $\psi_e = [\psi_{e+}, \psi_{e-}]^T$, where the spin-based Hamiltonian $H_s$ becomes $H_s = a_0\sigma_0 + a_1\sigma_1 + a_2\sigma_2 + a_3\sigma_3$ for the Pauli matrices $\sigma_{0,1,2,3}$ and their coefficients,

$$\begin{aligned}
a_0 &= \frac{ik \cdot \{k^2 - \omega^2 \cdot [\mu_0(\varepsilon_x + \varepsilon_y)/2 + \chi^2]\} + \omega^2 \chi \cdot \chi'}{2 \cdot (k^2 - \omega^2 \chi^2)} \\
a_1 &= -\frac{ik\omega^2\mu_0 \cdot (\varepsilon_x - \varepsilon_y)/2}{2 \cdot (k^2 - \omega^2 \chi^2)} \\
a_2 &= -\frac{\omega^3 \chi \mu_0 \cdot (\varepsilon_x - \varepsilon_y)/2}{2 \cdot (k^2 - \omega^2 \chi^2)} \\
a_3 &= \frac{i\omega\chi \cdot \{k^2 + \omega^2 \cdot [\mu_0(\varepsilon_x + \varepsilon_y)/2 - \chi^2]\} - k\omega \cdot \chi'}{2 \cdot (k^2 - \omega^2 \chi^2)}
\end{aligned} \tag{S13}$$

Although Eq. (S13) provides the rigorous explanation for the light-matter interaction in nonmagnetic and electrically anisotropic chiral materials, we consider the simplified case to clarify the physical origin: the preserved sum of $\varepsilon_x$ and $\varepsilon_y$ along the z-axis ($\varepsilon_x = \varepsilon_o + \Delta\varepsilon(z)$, $\varepsilon_y = \varepsilon_o - \Delta\varepsilon(z)$, and $\chi = \chi_o + \Delta\chi(z)$ where $\varepsilon_o$ and $\chi_o$ are real). We can then set $k^2 = \omega^2 \cdot (\mu_0\varepsilon_o + \chi_o^2)$, and Eq. (S13) becomes

$$\begin{aligned}
a_0 &= \frac{-ik \cdot (2\chi_o \cdot \Delta\chi + \Delta\chi^2) + \chi \cdot \Delta\chi'}{2 \cdot (\mu_0\varepsilon_o - 2\chi_o \cdot \Delta\chi - \Delta\chi^2)} \\
a_1 &= -\frac{ik\mu_0 \cdot \Delta\varepsilon}{2 \cdot (\mu_0\varepsilon_o - 2\chi_o \cdot \Delta\chi - \Delta\chi^2)} \\
a_2 &= -\frac{\omega\chi\mu_0 \cdot \Delta\varepsilon}{2 \cdot (\mu_0\varepsilon_o - 2\chi_o \cdot \Delta\chi - \Delta\chi^2)} \\
a_3 &= \frac{i\omega\chi \cdot (2\mu_0\varepsilon_o - 2\chi_o \cdot \Delta\chi - \Delta\chi^2) - k \cdot \Delta\chi'/\omega}{2 \cdot (\mu_0\varepsilon_o - 2\chi_o \cdot \Delta\chi - \Delta\chi^2)}
\end{aligned} \tag{S14}$$

The condition of the constant optical chirality ($\Delta\chi = 0$) derives more explicit form of the coefficients

$$\begin{aligned}
a_0 &= 0 \\
a_1 &= -\frac{ik}{2} \cdot \left(\frac{\Delta\varepsilon}{\varepsilon_o}\right) \\
a_2 &= -\frac{\omega\chi_o}{2} \cdot \left(\frac{\Delta\varepsilon}{\varepsilon_o}\right) \\
a_3 &= i\omega\chi_o
\end{aligned} \tag{S15}$$

and the Hamiltonian $H_s$ becomes $H_s = a_1\sigma_1 + a_2\sigma_2 + a_3\sigma_3$ as shown in the main manuscript.

**References for Supplemental Material**